\documentstyle[proceedings]{crckapb} 

\makeatletter                                            
\@ifundefined{epsfbox}{\@input{epsf.sty}}{\relax}        
\def\plotone#1{\centering \leavevmode                    
\epsfxsize=\columnwidth \epsfbox{#1}}                    
\def\plotone_reduction#1#2{\centering \leavevmode        
\epsfxsize=#2\columnwidth \epsfbox{#1}}                  
\makeatother                                             

\begin{opening}
\title{3-D SPH SIMULATIONS OF 
RADIATIVELY COOLING 
MAGNETIZED JETS
}

\author{E.M. de Gouveia Dal Pino }
\author{A.H. Cerqueira}
\institute{University of S\~ao Paulo, Instituto Astron\^omico e 
Geof\'isico\\
           Av. Miguel St\'efano, 4200, 04301-904 S\~ao Paulo, SP, 
Brazil}

\end{opening}

\runningtitle{3-D SPH SIMULATIONS OF MAGNETIZED JETS}

\begin{document}



Considerable amount of numerical work on magnetized, adiabatic, 
light jets has been done to study extragalactic jets (see, e.g., 
[1] for a review). Only  recently, 
 these MHD studies have been extended to heavy, 
 radiatively 
cooling jets [e.g., 2, 3, 8, 9, 10].

In this work, 
we summarize the results of three-dimensional smooth particle
magnetohydrodynamics numerical simulations of supermagnetosonic,
radiatively cooling jets [3]. Two initial magnetic configurations (in
$\sim$ equipartition with the gas) are considered: (i) a helical and
(ii) a longitudinal field  which permeate both the jet and the ambient
medium [2]. 
These MHD simulations  have been
compared with a baseline of previous non-magnetic radiative cooling
and adiabatic calculation [e.g., 4, 5, 6, 7]. 
 We find that magnetic fields have important effects on the
dynamics and structure of radiative cooling jets, especially at the
head. 
Both magnetic field geometries
are able to improve jet collimation, although this effect is more
pronounced in the helical case. 
In both
magnetic configurations, the confining pressure of the
cocoon is able to excite MHD Kelvin-Helmholtz pinch
and kink
modes that drive low-amplitude internal shocks and wiggling 
along the beam (see Fig. 1).  
These
modes are however, inhibited by the presence of  radiatively cooling 
and
by increasing the density contrast, $\eta$, between the jet and the
ambient medium [4, 5, 6, 7].   
As a consequence, the weakness of the induced
shocks makes it doubtful that they could produce by themselves the
bright knots observed in the overdense, radiatively cooling 
protostellar jets.
Also we find that the
presence of a helical field (Fig. 1, bottom) suppresses the formation of the
clumpy structure which is found to develop at the head of $purely$
$hydrodynamical$ [3] jets by fragmentation of the cold shell of shocked
material. On the other hand, a cooling jet embedded in a longitudinal
magnetic field (Fig. 1, top) retains the clumpy morphology at 
its head.  This fragmented
structure resembles the knotty pattern commonly observed in HH objects
behind the bow shocks of protostellar jets. This suggests that a 
predominantly
helical magnetic field configuration is unlikely at
the jet head in those cases [2].

\begin{figure}[h]
\plotone_reduction{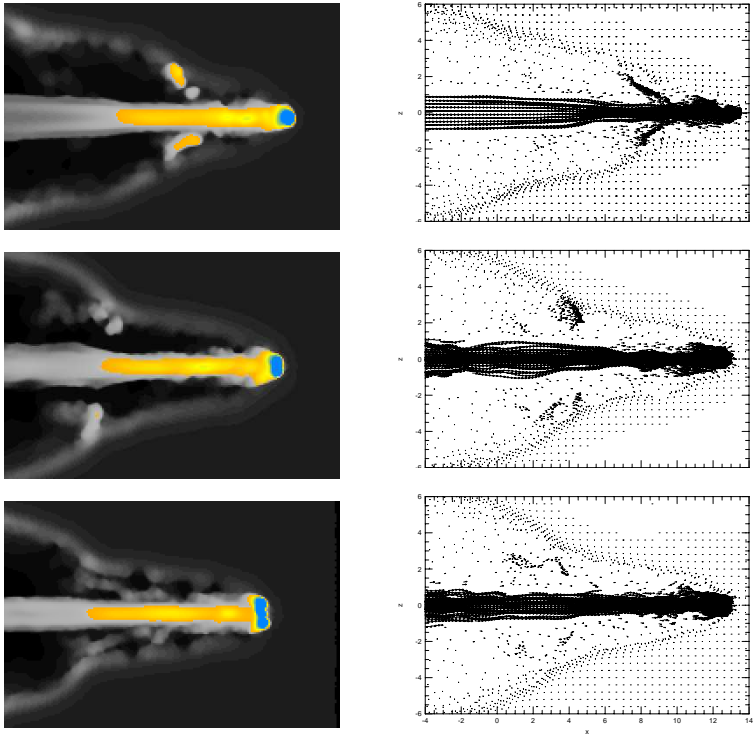}{.80}
\caption{Figure 1: Left $-$ gray-scale representation of the midplane
 density of the head of radiative cooling jets with $\eta=3$ and 
ambient Mach number $M_a = 24$: 
a hydrodynamical jet (HD3r, top), 
an MHD jet with initial longitudinal B-field (ML3r, middle), 
and an MHD jet with initial helical B-field (MH3r, bottom) 
after they have propagated $\approx 30 R_j$. 
Right $-$
mid-plane velocity field distribution [2].
}
\end{figure}

\end{document}